\documentclass{article}

\usepackage{arxiv}

\usepackage[utf8]{inputenc} % allow utf-8 input
\usepackage[T1]{fontenc}    % use 8-bit T1 fonts
\usepackage{hyperref}       % hyperlinks
\usepackage{url}            % simple URL typesetting
\usepackage{booktabs}       % professional-quality tables
\usepackage{amsfonts}       % blackboard math symbols
\usepackage{amsmath}
\usepackage{nicefrac}       % compact symbols for 1/2, etc.
\usepackage{microtype}      % microtypography
\usepackage{cleveref}       % smart cross-referencing
\usepackage{lipsum}         % Can be removed after putting your text content
\usepackage{graphicx}
\usepackage{doi}

\title{Diquarks and $\Lambda^0/\pi^+$, $\Xi^-/\pi^+$ ratios in the framework of the EPNJL model}

\newif\ifuniqueAffiliation
\uniqueAffiliationtrue

\ifuniqueAffiliation % Standard variant of author block
\author{ \hspace{1mm} A. V. Friesen  \\
	Joint Institute for Nuclear Research\\
	Dubna, Russia, 141980\\
	\texttt{avfriesen@theor.jinr.ru} \\
	\And
	{Yu. L. Kalinovsky} \\
	Joint Institute for Nuclear Research\\
	Dubna, Russia, 141980\\
	\texttt{kalinov@jinr.ru} \\
}
\else
\usepackage{authblk}

\author[1]{{A. V. Friesen\thanks{\texttt{avfriesen@theor.jinr.ru}}}%
}
\author[1]{{\hspace{1mm} Yu. L. Kalinovsky \thanks{\texttt{kalinov@jinr.ru}}}%
}
\affil[1]{Joint Institute for Nuclear Research, Dubna, Russia, 141980}

\fi

\renewcommand{\shorttitle}{Diquarks and $\Lambda^0/\pi^+$, $\Xi^-/\pi^+$ ratios }

\hypersetup{
pdftitle={Diquarks and $\Lambda^0/\pi^+$, $\Xi^-/\pi^+$ ratios in the framework of the EPNJL model},
pdfsubject={},
pdfauthor={A. V. Friesen, Yu. L. Kalinovsky},
pdfkeywords={PNJL model; baryon structure; diquark-quark bounding state, the hadronic deconfinement, baryon-to-meson ratio},
}

\begin{document}
\maketitle

\begin{abstract}
	The applicability of the effective models to the description of baryons and the behaviour of ratios of strange baryons to pions is discussed. In the framework of the EPNJL model, the Bethe - Salpeter equation is used to find masses of baryons, which are considered as diquark-quark state. Baryon melting is discussed at a finite chemical potential and a flavor dependence of the hadronic deconfinement temperature is pointed. It is shown that the description of the diquark-quark state at finite chemical potential is limited due to the occurrence of the Bose condensate. This effect is strongly manifested in the description of light diquarks and baryons. Both  $\Lambda^0/\pi^+$ and $\Xi^-/\pi^+$ ratios show a sharp behaviour as functions of $T/\mu_B$ variable, where T and $\mu_B$ are calculated along the melting lines.
\end{abstract}

% keywords can be removed

\section{Introduction}
In our previous works \cite{Friesen:2019azp,Blaschke:2021yml,Blaschke:2020bzh,Friesen:2014mha} the peak-like structure in a $K^+/\pi^+$ ratio was discussed in the framework of the Polyakov-loop extended Nambu-Jona-Lasinio  model (PNJL) and its modifications including the vector interaction. The interest in this structure is due to the search for signals of a phase transition from hadron phase to the quark-gluon plasma (QGP) formation during the heavy ion collision \cite{NA49:2002pzu,Andronic:2008gu}. The quick rise in the $K^+/\pi^+$ ratio is associated with the phase transition in the medium, while the jump from the maximum value to the constant valley is explained as the QGP formation during the collision. This is a consequence of the fact that  after the deconfinement transition occurs in the system the strangeness yield becomes independent of the collision energy \cite{Palmese:2016rtq,Gazdzicki:1998vd,Cohen:1991nk,Nayak:2010uq}. Recent investigations showed that the $K^+/\pi^+$ peak strongly depends on the volume of the system and tends to be less pronounced in small-size systems \cite{Palmese:2016rtq,Lewicki:2020mqr}.

The meson-to-meson ratios are widely considered both in theoretical and experimental works, in contrast to the baryon - to - meson ratios although they also have a peak-like structure. In the work \cite{Oeschler:2017bwk} in the framework of the thermal model it was shown, that unlike to $K^+/\pi^+$-ratio, the peak for $\Lambda^0/\pi^+$ does not disappear with reducing of the system size. 

The choice of the (E)PNJL model for such investigations is conditioned by the possibility to describe within the model both the chiral phase transition and the deconfinement transition, which can give a hint for understanding the nature of peaks at least quantitatively. For the next step it is interesting consider baryon - to - meson ratios in the framework of the model. The controversy of the applicability and the complexity of this task are related to the problem of  describing baryons in the frame of the NJL-like models. The most detailed and exact description of baryons requires solving of the three-body Faddeev equation. Which leads to considering of baryons as bound state of a quark and diquark \cite{Buck:1992wz,Ebert:1996ab}. The so-called “static approximation” of Faddeev equation leads to the Bethe-Salpeter equation, which is based on the polarisation loop in the diquark-quark scattering channel \cite{Blanquier:2011zz}. But the diquark-quark structure of baryons leads to the non-obviousness of the description of baryons in a dense medium, where the formation of a Bose condensate occurs and the diquark states melt.

The results of our calculations and discussion about aspects of applicability of the model are presented in the last section of the article.

\section{SU(3) PNJL Lagrangian}

The complete Lagrangian of the SU(3) PNJL model with the vector interaction and $U_A(1)$ anomaly has the form \cite{Vogl:1991qt,Blanquier:2011zz}:  
\begin{eqnarray}
\mathcal{L\,} & = & \bar{q}\,(\,i\,{\gamma}^{\mu}\,D_{\mu}\,-\,\hat
{m} - \gamma_0\mu)\,q  \nonumber \\
&+& \frac{1}{2}\,g_{S}\,\,\sum_{a=0}^{8}\,[\,{(\,\bar{q}\,\lambda
^{a}\,q\,)}^{2}\,\,+\,\,{(\,\bar{q}\,i\,\gamma_{5}\,\lambda^{a}\,q\,)}%
^{2}\,]  \nonumber \\
&-& \frac{1}{2}g_{\rm V} \sum_{a=0}^{8}\,\,[(\bar{q}\gamma_\mu\lambda
^{a} q)^{2} + (\bar{q}\gamma_\mu i \gamma_5\lambda^{a}q )^{2}] \nonumber \\
&-&  \sum_{\alpha} g_{\rm diq}^\alpha\sum_{i,j} \left(\bar{q}_a\Gamma^i_\alpha q^C_b \right) \left(\bar{q}^C_d\Gamma^j_\alpha q_e \right) \varepsilon^{abc}\varepsilon^{de}_c\nonumber \\
&+& \mathcal{L}_{\rm det} -  \mathcal{U}(\Phi, \bar{\Phi}; T),
\label{lagr}%
\end{eqnarray}  
where $q=(u,d,s)$ is the quark field with three flavours, ${q}^C$ is the charge conjugated quark field, $\hat{m}=\mbox{diag}(m_{u},m_{d},m_{s})$ is the current quark mass matrix, $g_{\rm S}$, $g_{\rm V}$, $g_{\rm diq}$ are the coupling constants. The entanglement PNJL model (EPNJL) includes the constants $g_{\rm S}, g_{\rm V}$ introduced as functions of T to enhance the coupling between quarks and the gauge field \cite{Sugano:2014pxa, Blaschke:2021yml}.  $\Gamma^j_\alpha$ is a product of Dirac matrices $\gamma^\mu$ and Gell-Mann matrices  $\lambda^{\alpha}$, where index $\alpha$ describes the type of diquarks. The covariant derivative is D$_\mu = \partial^\mu -i A^\mu$, where  $A^\mu$   is the gauge field with $A^0= -  iA_4$ and $A^\mu(x) = g_SA^\mu_a\frac{\lambda_a}{2}$ absorbs the strong interaction coupling. The Kobayashi - Masakawa - t'Hooft (KMT) interaction is described by the term 
$$\mathcal{L}_{\rm det} = g_D \,\,\{\mbox{det}\,[\bar{q}\,(\,1\,+\,\gamma_{5}\,)\,q\,]+\mbox{det}\,[\bar{q}\,(\,1\,-\,\gamma_{5}\,)\,q\,]\,\}$$

The last term is the effective potential $\mathcal{U}(\Phi, \bar{\Phi}; T)$, expressed in terms of the traced Polyakov loop $\Phi = N_c^{-1}{\rm tr}_c \langle L(\bar{x})\rangle$ \cite{Ratti:2005jh}, where
 
\begin{equation}
     L(\bar{x}) = \mathcal{P}{\rm exp}\left[\int_0^{\infty}d\tau A_4(\bar{x},\tau)\right].
\end{equation}

The effective potential describes the confinement properties (Z$_3$-symmetry) and is constructed on the basis of  Lattice inputs in the pure gauge sector. In this work, we use the standard polynomial form of the effective potential  \cite{Blanquier:2011zz,Friesen:2014mha}. The effect of the vector interaction on the position of the critical end point in the phase diagram and on the behaviour of the peak in the $K^+/\pi^+$ ratio was discussed in previous works \cite{Friesen:2019azp,Blaschke:2021yml,Blaschke:2020bzh,Friesen:2014mha}.

The grand potential density $\Omega(T,\mu_i)$ in the mean-field approximation with $g_{\rm V} = 0$ can be obtained from the Lagrangian density (\ref{lagr}) and leads to a set of self-consistent equations:
\begin{equation}
\frac{\partial \Omega}{\partial \langle\bar{q_i}q_i\rangle}=0, \ \ \frac{\partial \Omega}{\partial \Phi}=0, \ \ \frac{\partial \Omega}{\partial \bar{\Phi}}=0, 
\end{equation}
where $\Phi, \bar{\Phi}$ are the Polyakov fields. The gap equations for quark masses are: 
\begin{eqnarray}
m_i &=& m_{0i} - 2 g_S\langle\bar{q_i}q_i\rangle - 2g_D\langle\bar{q_j}q_j\rangle\langle\bar{q_k}q_k\rangle, \label{eqgap} 
\end{eqnarray}
where $i, j, k = u, d, s$ are chosen in cyclic order, $m_i$ are the constituent quark masses, the quark condensates are:
\begin{eqnarray}
\langle\bar{q_i}q_i\rangle &=& - 2 N_c \int\frac{d^3p}{(2\pi)^3}\frac{m_i}{E_i}(1 - f^+_\Phi(E_i) - f^-_\Phi(E_i))
\end{eqnarray}
with  modified Fermi functions $f^\pm_\Phi(E_i)$: 
\begin{eqnarray}
f^+_\Phi(E_f)&=&
\frac{(\bar{\Phi}+2{\Phi}Y)Y+Y^3}{1+3(\bar{\Phi}+{\Phi}Y)Y+Y^3}
~,\label{fPhi} \\
f^-_\Phi(E_f)&=&
\frac{({\Phi}+2\bar{\Phi}\bar{Y})\bar{Y}+\bar{Y}^3}{1+3({\Phi}+\bar{\Phi}\bar{Y})\bar{Y}+\bar{Y}^3}
~, \label{fAPhi}
\end{eqnarray}
where $Y={\rm e}^{-(E_i-\mu_f)/T}$ and $\bar{Y}={\rm e}^{-(E_i+\tilde{\mu_i})/T}$. 

%%%%%%%%%%%%%%%%%%%%%%%%%%%%%%%%%%%%%%%%%%

\section{Mass equations for standard particles}

\begin{figure}[h]
\centerline{
\includegraphics[width = 0.5\columnwidth]{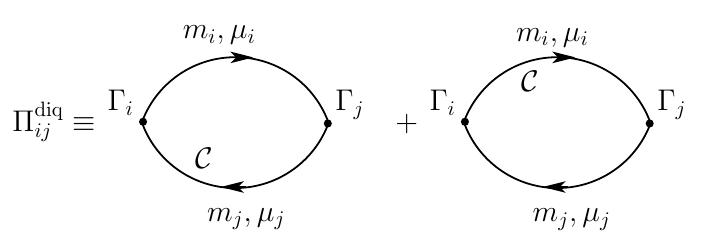}}
\caption{Polarization loops for diquarks, "$\mathcal{C}$" is the charge conjugation operator.}
\label{loop_diq}
\end{figure}

To describe mesons and diquarks as quark-antiquark and quark-quark bound states, the random-phase approximation is usually used  in the framework of the (E)PNJL model. The masses of bound states are defined by the polarization function $\Pi_{ij}$. 
\begin{figure}[h]
\centerline{
\includegraphics[width = 4cm]{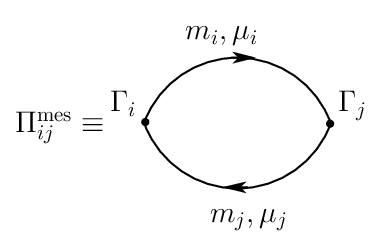}}
\caption{Polarization loop for mesons.}
\label{loop_mes}
\end{figure}
Mesons as a quark-antiquark system have polarization loop shown in Fig.\ref{loop_mes}. The polarization function for mesons is defined as
\begin{equation}
\Pi_{ij} = \int \frac{d p}{(2\pi)^4} {\rm tr}\lbrace  S^{i}(\hat{q_i}, m_i)\Gamma_j S^{j}(\hat{q_j}, m_j) \Gamma_i\rbrace, 
\end{equation}
where $\Gamma_{i,j}$ are the vertex matrices (Fig.\ref{loop_diq}) and $S^i(\hat{q_i}, m_i) = ( \hat{q_i}+\gamma_0 (\mu_i - i A_4) - m_i)^{-1}$ is the $i-$ flavour  quark propagator. The meson mass is obtained from the Bethe-Salpeter equation in the meson rest frame ($\bar{P} = 0$) 
\begin{equation}
1-P_{ij}\Pi_{ij}(P_{0}=M,\bar{P}=0)=0~, \label{rdisp}
\end{equation}
where the function $P_{ij}$ depends on the type of meson (see details in \cite{Rehberg:1995kh}). For example, for the pion, which is a pseudo-scalar meson $P_{ud}=g_{S}+g_{D}\left\langle\bar{q}_{s}q_{s}\right\rangle$.

Diquarks are considered as two-quark system and to describe the polarization loop in the same way, the ''antiquark'' is replaced by its charge conjugate propagator. Then two diagrams should be taken into account, however it can be shown that they give the same result. Polarization loops for diquarks are shown in Fig. \ref{loop_diq},  where $\mathcal{C} = i \gamma_0\gamma_2$ is the charge conjugation operator, $\Gamma_{i, j}$ are the vertex functions. 

According to group theory, diquarks can be represented by symmetric and antisymmetric wave functions both in  colour and flavour spaces. Since  diquarks are used to construct  baryons which are ''white objects'',  only diquarks with a colour antisymmetric wave function are considered.  According to the interaction type, diquarks can be of scalar, pseudo-scalar, axial and vectorial types following the rule that the diquark wave function is total antisymmetric (see Table \ref{tab_diq}).

\begin{table}
\caption{List of mesons and diquarks.\label{tab_diq}}
\centering
\begin{tabular}{ccccc}
\toprule
$\Gamma$   & Meson type & Possible mesons & Diq. type & Possible diq.\\
\midrule
i$\gamma_5$  & pseudoscalar & $\pi$, K & scalar & \\
$1$  & scalar & $\sigma$, $K_0^*$ & pseudoscalar & $(ud)$, $(us)$, $(ds)$\\
$\gamma^\mu i \gamma_5$  & axial-vector & $a_1^*$, $K_1^*$ & vector & \\
\midrule
i$\gamma^\mu$  & vector & $\rho$, $K^*$ & axial-vector & $[ud]$,$[us]$,$[ds]$,$[uu]$,$[dd]$,$[ss]$ \\
\bottomrule
\end{tabular}
%\noindent{\footnotesize{\textsuperscript{1} Tables may have a footer.}}
\end{table}

The Bethe-Salpeter equation for diquarks in the rest frame is 
\begin{equation}
    1- Z_{\rm diq}\Pi_{ij}(P_{0}=M_{\rm diq},\bar{P}=0) = 0,
\end{equation}
with polarization operators corresponding to Fig.\ref{loop_diq}
\begin{eqnarray}
\Pi_{ij}^{(1)} &=& \int \frac{d p}{(2\pi)^4} tr\lbrace  S^{i}(\hat{q_i}, m_i)\Gamma_j S^{j C}(\hat{q_j}, m_j) \Gamma_i \rbrace, \\
\Pi_{ij}^{(2)} &=& \int \frac{d p}{(2\pi)^4} tr\lbrace  S^{i C}(\hat{q_i}, m_i)\Gamma_j S^{j}(\hat{q_j}, m_j) \Gamma_i \rbrace, 
\end{eqnarray}
which give the same result, $S^{iC}(\hat{q_i} ) = ( \hat{q_i} - \gamma_0 (\mu_i+ i A_4) - m_i)^{-1}$ is the propagator of the charge conjugated quark and $Z_{\rm diq}$ is the coupling constant for diquarks. $Z_{\rm diq} =  g^s_{\rm diq}$  for scalar and pseudoscalar diquarks and $Z_{\rm diq} = g^s_{\rm diq}/4$ for vector and axial-vector diquarks. According to the Lagrangian Eq. (\ref{lagr}) and the Fierz transformation, the coupling constant $g^s_{\rm diq}$ is referred to $g_{\rm S}$ as $g^s_{\rm diq} = 3g_{\rm S}/4$ and is usually chosen as $g^{\rm s}_{\rm{diq}} \sim (0.705 - 0.75) g_{\rm S}$ \cite{Vogl:1991qt,Blanquier:2011zz}.  

%================================================
The description of baryons is a more complicated task, since they are a complex structures of three quarks, coupled through the exchange of gluons. Thus, the modelling of a three-body system is required and the Faddeev equation has to be considered. However some simplifications of the Faddeev equation allow to consider the baryon as a diquark-quark bound state \cite{Vogl:1991qt,Buck:1992wz,Ebert:1996ab,Blanquier:2011zz}. Considering the static approximation for the four-point interaction leads to the loop structure of the transition matrix and the matrix Bethe-Salpeter-like equation for the baryon mass:
\begin{equation}
1 - \Pi_{i(D)}(k_0,\vec{k})\cdot Z_{ij} = 0,
\label{BS_bar}
\end{equation}
where the constant $Z_{ij}$ is defined as 
\begin{equation}
Z_{ij} = \frac{g_{ik}g_{jk}}{m_{k}},
\label{Z_bar}
\end{equation}
where $g_{ij}$ is the diquark-quark coupling $D_{ij}\rightarrow q_iq_j$ and $g_{ud}$ includes the factor (-2) \cite{Blanquier:2011zz}. In the same way as for diquarks, two diquark-quark loops should be considered, however as shown in \cite{Blanquier:2011zz} they give the same result.
\begin{figure}[h]
\centerline{
\includegraphics[width = 0.5\columnwidth]{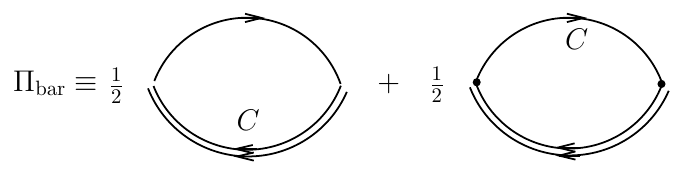}}
\caption{The baryon loops function.}
\label{loop_bar}
\end{figure}

Just as for diquarks, two loops should be taken into account, and it can be easily  shown that they give the same results
\begin{eqnarray}
\Pi_{i(D)}^{(1)} &=& \int \frac{d p}{(2\pi)^4} tr\lbrace  S^{i}(\hat{q_i}, m_i)\Gamma_j S_D^{j C}(\hat{q_j}, m_j) \Gamma_i \rbrace, \\ \label{Pol_bar1}
\Pi_{i(D)}^{(2)} &=& \int \frac{d p}{(2\pi)^4}  tr\lbrace  S^{i C}(\hat{q_i}, m_i)\Gamma_j S_D^j(\hat{q_j}, m_j) \Gamma_i \rbrace.
\label{Pol_bar2}
\end{eqnarray}

It should be noted here that  the axial-diquark contribution to the members of the baryon octet is neglected \cite{Blanquier:2011zz,Torres-Rincon:2015rma}.

%%%%%%%%%%%%%%%%%%%%%%%%%%%%%%%%%%%%%%%%%%

\section{Numerical results}

In previous works \cite{Friesen:2019azp,Blaschke:2021yml,Blaschke:2020bzh}
a detailed study of $K/\pi$ ratios was carried out in the framework of PNJL-like models. As the collision energy $\sqrt{s_{\rm NN}}$ never appears in effective models, a trick with fitting $\sqrt{s_{\rm NN}}$ by the pair  $(T, \mu_B)$  from the statistical model was used. In the statistical model, the temperature and the baryon chemical potential of freeze-out are assigned to each collision energy (e.g., as suggested by Cleymans et al. \cite{Oeschler:2017bwk}). Supposing that the chiral phase transition line in the EPNJL model corresponds to the freeze-out, the K/$\pi$ ratio can be considered as a function of a new variable $T /\mu_B$ instead of $\sqrt{s_{\rm NN}}$ , where $(T, \mu_B)$ are taken along the phase transition line.

The phase diagram has a classic structure with smooth crossover at low chemical potentials and the first order chiral phase transition at high chemical potential. The PNJL model has a crossover temperature $T_c = 0.27$ GeV higher than the Lattice prediction $T_c\sim 0.17$ GeV. An extended version of the PNJL model (EPNJL) with $f_{V}(T)$, $g_{S}(T)$ was introduced to reduce the critical temperature of the crossover to lower value $T^{\rm EPNJL}_c = 0.18$ GeV due to enhanced interaction between quarks and gauge sector \cite{Friesen:2014mha}. For more detailed study, the meson masses were calculated both in the Bethe-Salpeter and the Beth-Uhlenbeck approaches, the latter is preferable for considering mesons in hot and dense matter, since it takes into account their spectral functions and correlations.

For the effective models, the ratio of the particle number can be calculated in terms of the ratio of the number densities:
\begin{eqnarray}
n = d\int_0^\infty p^2dp\frac{1}{e^{\beta(\sqrt{p^2+m^2}\mp\mu)}\pm1 },
\end{eqnarray}
where $d$ is the corresponding degeneracy factor, upper sign in denominator refers to fermions and lower refers to bosons and $\beta = T^{-1}$. The pion chemical potential is a phenomenological parameter and it was chosen as a constant, the baryon chemical potential is calculated as the sum of chemical potentials of constituent quarks. The degeneracy factors are calculated as $(2s +1)(2I+1)$, for $\Lambda^0$ is 2 and for $\Xi^{-}$ is 4. 

\begin{figure}[h]
\centerline{
\includegraphics[width = 6cm]{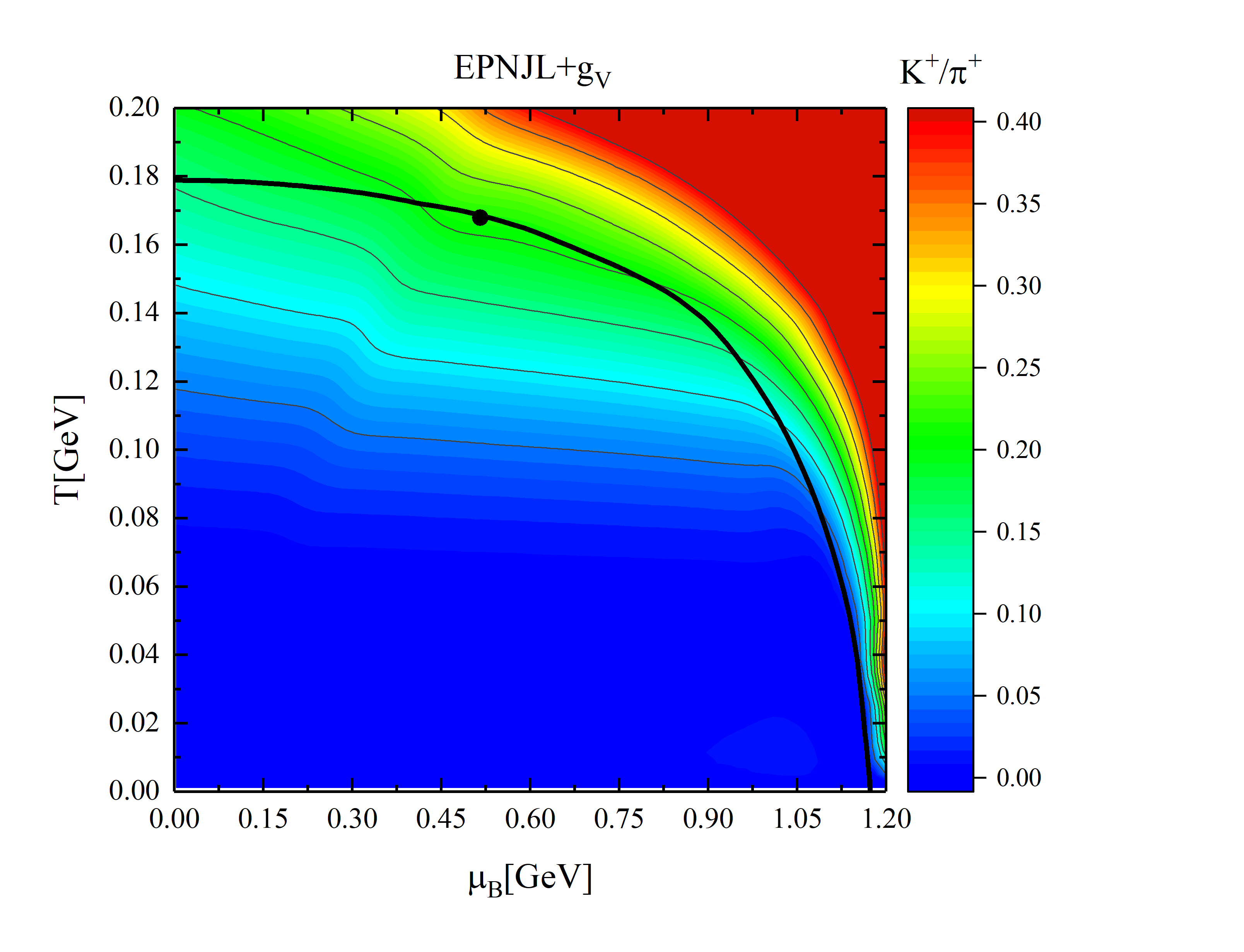}
\includegraphics[width = 6cm]{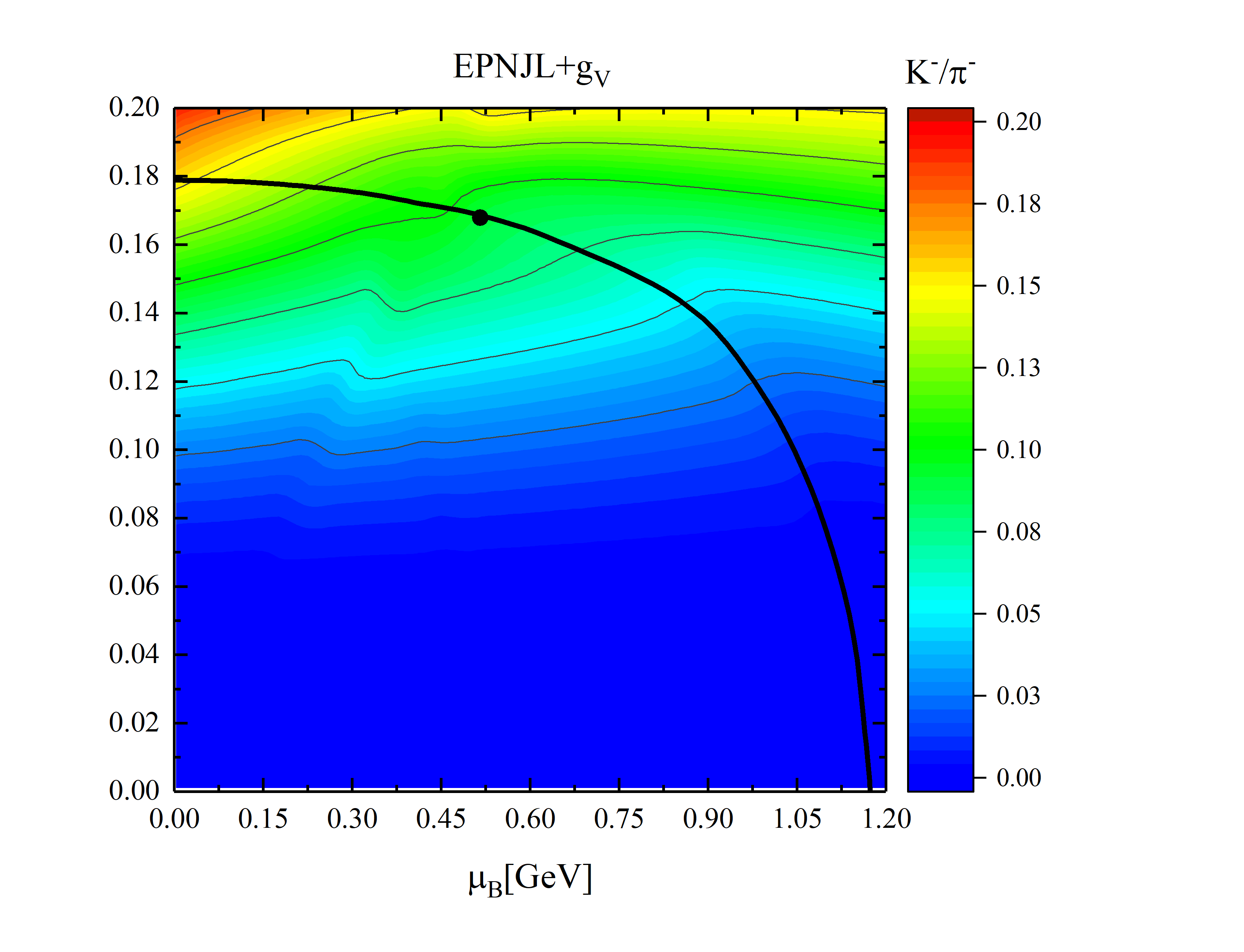}
}
\caption{$K^+/\pi^+$ (left) and $K^{-}/\pi^{-}$ (right) on the $T-\mu_B$ plane for the EPNJL model with $g_{\rm V} = 0.6 g_{\rm S} $ (no CEP) and $\mu_\pi = 0.147$ GeV.  The black dot indicates the maximum of $K^+/\pi^+$ ratio on the line of pseudocritical temperatures for the chiral transition (our proxy for chemical freeze out).}
\label{Kpi_contour}
\end{figure}

The Fig.\ref{Kpi_contour} shows contour graphs for $K^+/\pi^+$ and $K^-/\pi^-$ ratios obtained in the Beth-Uhlenbeck approach of the EPNJL model with $g_{V} = 0.6 g_{S}$ \cite{Blaschke:2020bzh}. The black lines show the phase transition (corossover) lines. It can be seen, that when shifting along the phase transition line from low to high temperature,  the trajectory shows a quick enhancement and then a fall for  $K^+/\pi^+$ and a smooth increase for the  $K^-/\pi^-$.

The results in Fig.\ref{Kpi_contour} are presented for the case with fixed pion chemical potential $\mu_\pi = 0.147.6$ GeV. In order to reproduce the experimental data, the dependence of pion and the strange quark chemical potentials on the variable $x=T/\mu_B$ should be introduced.  The expressions should describe the increase in the pion chemical potential with $x$ and the decrease in the strange quark chemical potential, respectively. For their $x$- dependence the functions of the Woods-Saxon form is suggested \cite{Blaschke:2021yml}
\begin{eqnarray}
\mu_\pi(x) &=& \mu_\pi^{\mathrm{min}}  + \frac{\mu_\pi^{\mathrm{max}} - \mu_\pi^{\mathrm{min}}}{1 + \exp(-(x - x_\pi^{\mathrm{th}})/\Delta x_\pi)) }, \label{eq:mupi}
%\nonumber
\\
\mu_s(x) &=&  \frac{\mu_s^{\mathrm{max}}}{1 + \exp(-(x - x_s^{\mathrm{th}})/\Delta x_s)) } .
\label{eq:mus}
\end{eqnarray}
Parameters in Eqs.(\ref{eq:mupi}), (\ref{eq:mus}) were obtained from fitting the experimental data (see for details \cite{Blaschke:2020bzh,Blaschke:2021yml}). The best parameter values  for the EPNJL model are 
$\mu_\pi^{\mathrm{max}}= 107\pm10$ MeV, 
$\mu_\pi^{\mathrm{min}}=92$ MeV,
$x_\pi^{\mathrm{th}}=0.409$, 
$\Delta x_\pi = 0.00685$.
And for $\mu_s$ parameter values are
$\mu_s^{\mathrm{max}}/\mu_u^{\mathrm{crit}}= 0.205$, 
$x_s^{\mathrm{th}}=0.223$, $\Delta x_s = 0.06$.

\begin{figure}[!h]
\centerline{
\includegraphics[width = 0.45\columnwidth]{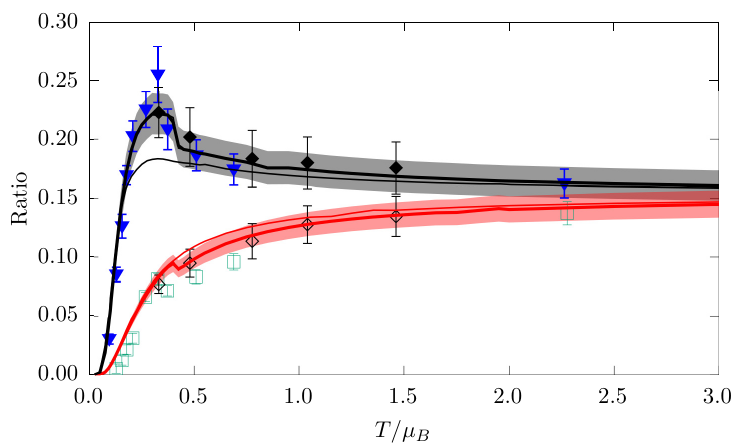}
\includegraphics[width = 0.45\columnwidth]{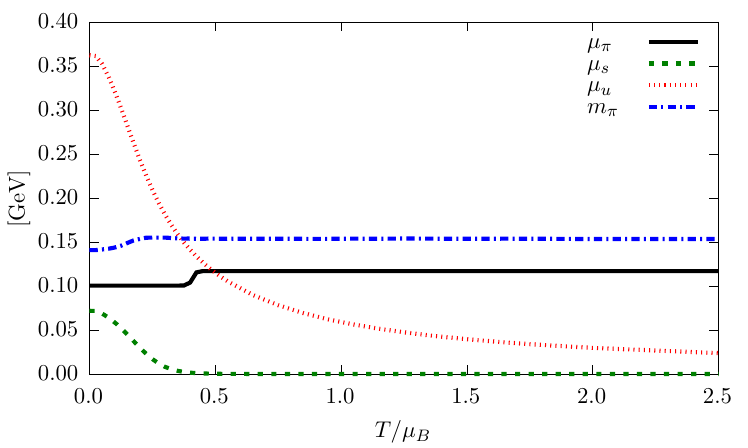}
}
	\caption {Left panel: The  $K^+/\pi^+$ (black lines) and $K^-/\pi^-$ (red lines) ratios are shown as function of $T/\mu_B$. Thin lines correspond to the case when $\mu_s=0$ and fixed $\mu_\pi= 0.147$ GeV.
Right panel: chemical potentials and pion mass as functions of $T/\mu_B$.
}
	\label{nKnpi_EPNJLwithx}
\end{figure}

In the left panel of Fig.\ref{nKnpi_EPNJLwithx}  $K^+/\pi^+$ (black lines) and $K^-/\pi^-$ (red lines) are shown as  functions of $T/\mu_B$ obtained in the Beth-Uhlenbeck approach for the EPNJL model with $g_{\rm V} = 0$ and $\mu_s$ and $\mu_\pi$ calculated according to Eqs.(\ref{eq:mupi}, \ref{eq:mus}). Thin lines correspond to the case when $\mu_s=0$ and fixed $\mu_\pi$. The shaded region corresponds to the error band due to normalization to high $x$ RHIC and LHC data. The behaviour of potentials is shown in right panel of Fig.\ref{nKnpi_EPNJLwithx}. 

Figures \ref{Kpi_contour} and \ref{nKnpi_EPNJLwithx}  demonstrate that the "horn" structure in the $K^+/\pi^+$ ratio is less sensitive to the structure of the phase diagram and more sensitive to the properties of the medium. At $g_{\rm V} = 0.6 g_{\rm S}$ the phase diagram has a smooth crossover transition at high density instead of the first order transition when $g_{\rm V} = 0$. Nevertheless the ratio keeps a ''horn'' structure. Changing the matter properties by modelling the chemical potentials for pions and $s$-quark leads to the possibility of reproducing  the experimental data.

%========================== baryons ===========================
The present work is devoted to the description of baryons and $\Xi^-/\pi$, $\Lambda^0/\pi$ ratios within this kind of models. According to Eqs.(\ref{BS_bar}), the Bethe-Salpeter equation for barions has a matrix form ${\rm det}(1-Z\Pi) = 0$ where for $\Lambda$:
\[
\Pi^{\Lambda} = 
\begin{bmatrix}
    \Pi_{(ds)u}       & 0 & 0 \\
    0      & \Pi_{(us)d} &0 \\
    0       & 0 & \Pi_{(ud)s}
\end{bmatrix},
\ \ \ \ \ \ 
Z^{\Lambda} = 
\begin{bmatrix}
    0  & Z_{ud} & Z_{us} \\
    Z_{du}      & 0 & Z_{ds} \\
    Z_{su}       & Z_{sd} & 0
    \end{bmatrix}
\]
and for $\Xi$
\begin{equation}
 \Pi^\Xi = \Pi_{(us)s}, \ \ \ \ \ \ Z^{\Xi} = Z_{ds},  
\end{equation}
functions $\Pi$, $Z$ are presented in Eqs. (\ref{Z_bar}-\ref{Pol_bar2}).

%\begin{eqnarray}
%1 &-& 2 \Pi_{u(ds)}\Pi_{d(us)}\Pi_{s(ud)}Z^{ud}Z^{us}Z^{ds} - %\Pi_{u(ds)}\Pi_{d(us)}(Z^{ud})^2 \nonumber \\
%&-& \Pi_{u(ds)}\Pi_{s(ud)}(Z^{ud})^2 - \Pi_{d(us)}\Pi_{s(ud)}(Z^{ds})^2 = 0,
%\end{eqnarray}
%where $\Pi, Z$ are described in Eqs.(\ref{Z_bar}, \ref{Pol_bar}). 

The calculations were performed with the parameter set $m_{u0}=m_{d0} = 4.75$ MeV, $m_{s0}= 0.147$ GeV, $\Lambda = 0.708$ GeV, $g_{S}\Lambda^{2} = 1.922$, $g_{D}\Lambda^5 = 10.0$, $g_{\rm V} = 0$, $g_{\rm diq} = 0.725 g_{\rm S}$.  The choice of the parameter set was driven by the requirement to have the proton and $\Lambda$ masses below the threshold $M_D + m_q$.

The  dissociation temperature for baryons is postulated from their diquark-quark structure. The Mott temperature ($T^{\rm bar}_{\rm Mott}$) is a temperature for which  the mass of baryons is equal to the sum of quark and diquark masses   \cite{Blanquier:2011zz,Torres-Rincon:2015rma,Wang:2010iu}. To avoid the situation when the diquark melts at a lower temperature and the baryon still exists, the baryon deconfinement temperature is chosen as
$$T^{\rm bar}_{\rm dec} = {\rm min}\lbrace T^{\rm bar}_{\rm Mott}, T^{\rm diq}_{\rm Mott}\rbrace.$$

Nevertheless, even if the  diquark becomes already unbound due to the Mott effect, the baryon can still be  bound as a three-particle state (so-called "borromean state") \cite{Blaschke:2015sla, Wang:2010iu}. In this case the ''dissociation'' temperature for baryons should be considered as temperature when the baryon melts into 3 quarks ($T_{\rm diss}$). 
\begin{figure}[h]
\centerline{
\includegraphics[width = 0.5\columnwidth]{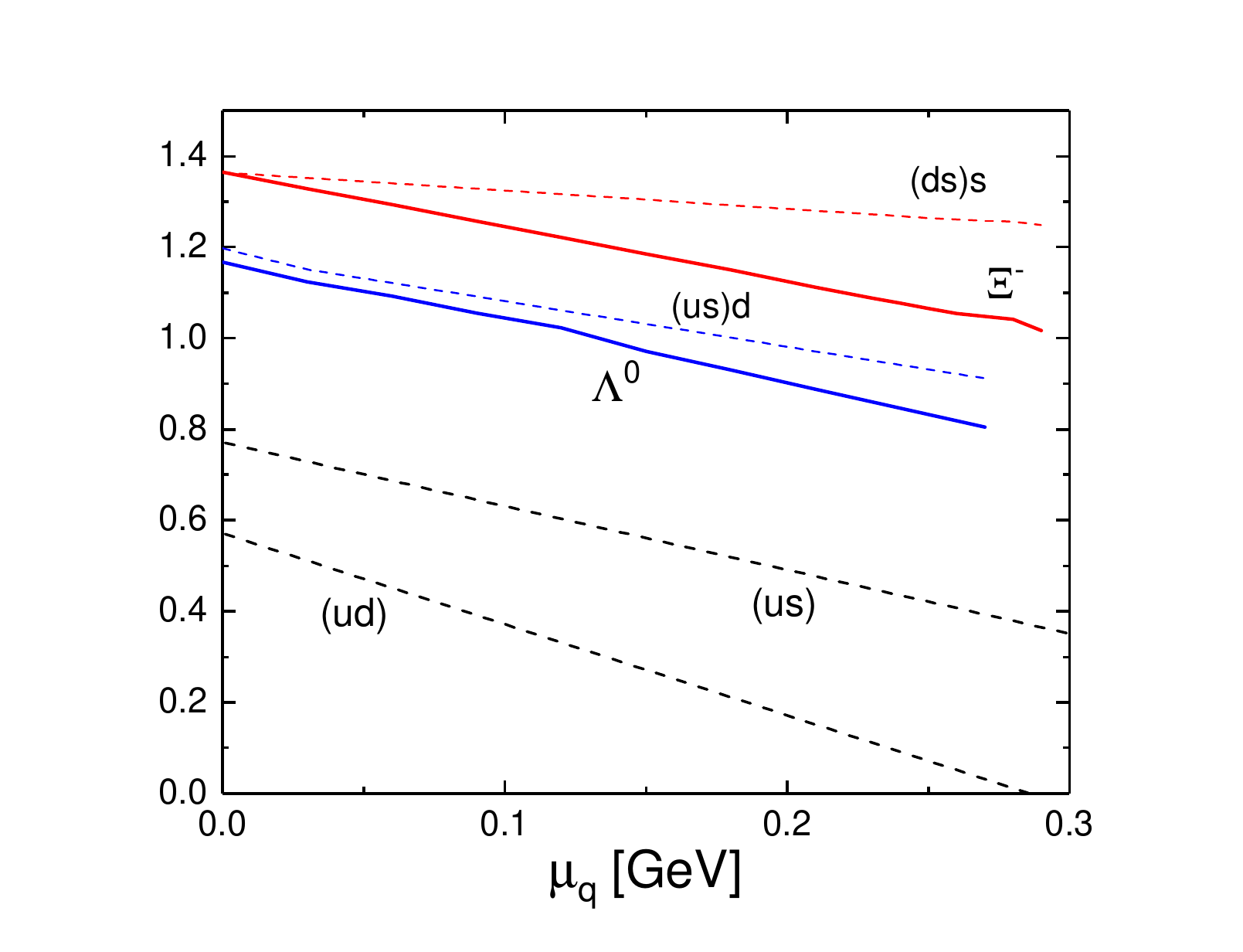}
\includegraphics[width = 0.5\columnwidth]{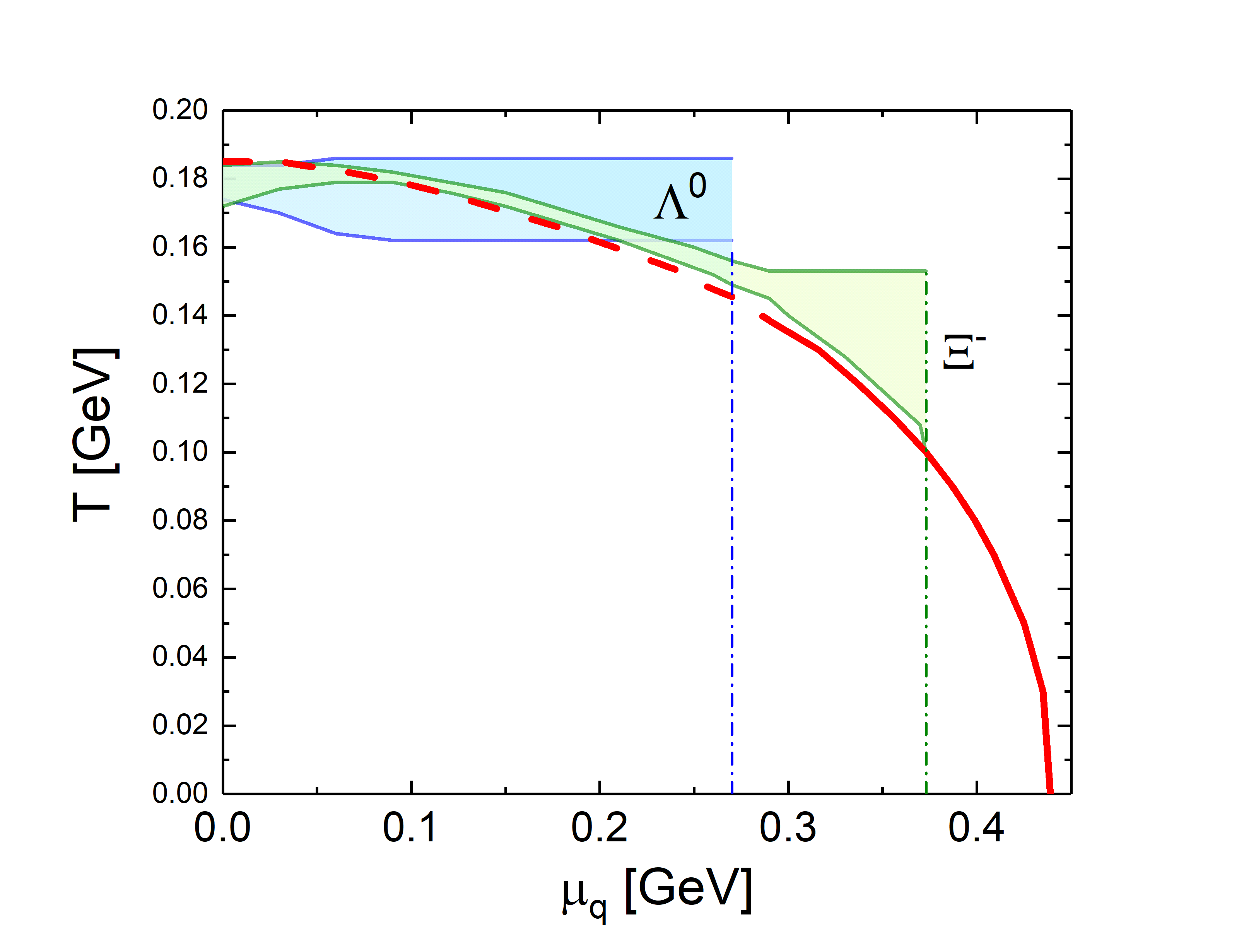}}
\caption{Left: masses of diquarks, baryons and their threshold $M_D + m_q$ as functions of $\mu_q$. Right: phase diagram of the EPNJL model (red line). The shaded areas show the borders $\lbrace T^{\rm bar}_{\rm dec}, T_{\rm diss}\rbrace$ for baryons: light blue for $\Lambda$ and light-green for $\Xi$.  }
\label{phase_diagr}
\end{figure}

The dissociation boundaries of baryons corresponding to $\lbrace T^{\rm bar}_{\rm dec}, T_{\rm diss}\rbrace$ are shown in Fig.\ref{phase_diagr} (right panel) with the
 light-blue shaded area for $\Lambda$ and the light-green one for $\Xi$. The red line corresponds to the phase diagram of the EPNJL model with $g_{\rm V} = 0$: dashed line corresponds to the crossover and solid line corresponds to the first order transition. As can be seen in the Fig.\ref{phase_diagr}, $\Xi^-$ baryon is described till $\mu_q\sim 0.37$ GeV, which is higher then $\mu_q \sim 0.27$ GeV for $\Lambda$. It appears from the fact that $\Xi^-$ is considered as combination of the scalar $(ds)$ diquark and s-quark, unlike $\Lambda$, which is a superposition of $(ud)+s$ and $(ds)+u$ ($(us)+d$) states. The diquark with heavy quark  survives at higher values of the chemical potential than light diquarks. The left panel of Fig. \ref{phase_diagr} shows masses of diquarks (dashed), baryons  (solid) and their thresholds $M_D + m_q$ (short-dashed) as functions of chemical potential $\mu_q$. Light diquarks melt at lower densities (or chemical potentials) due to the origin of the Bose-Einstein condensate. 
 
\begin{figure}[h]
\centerline{
\includegraphics[width = 0.5\columnwidth]{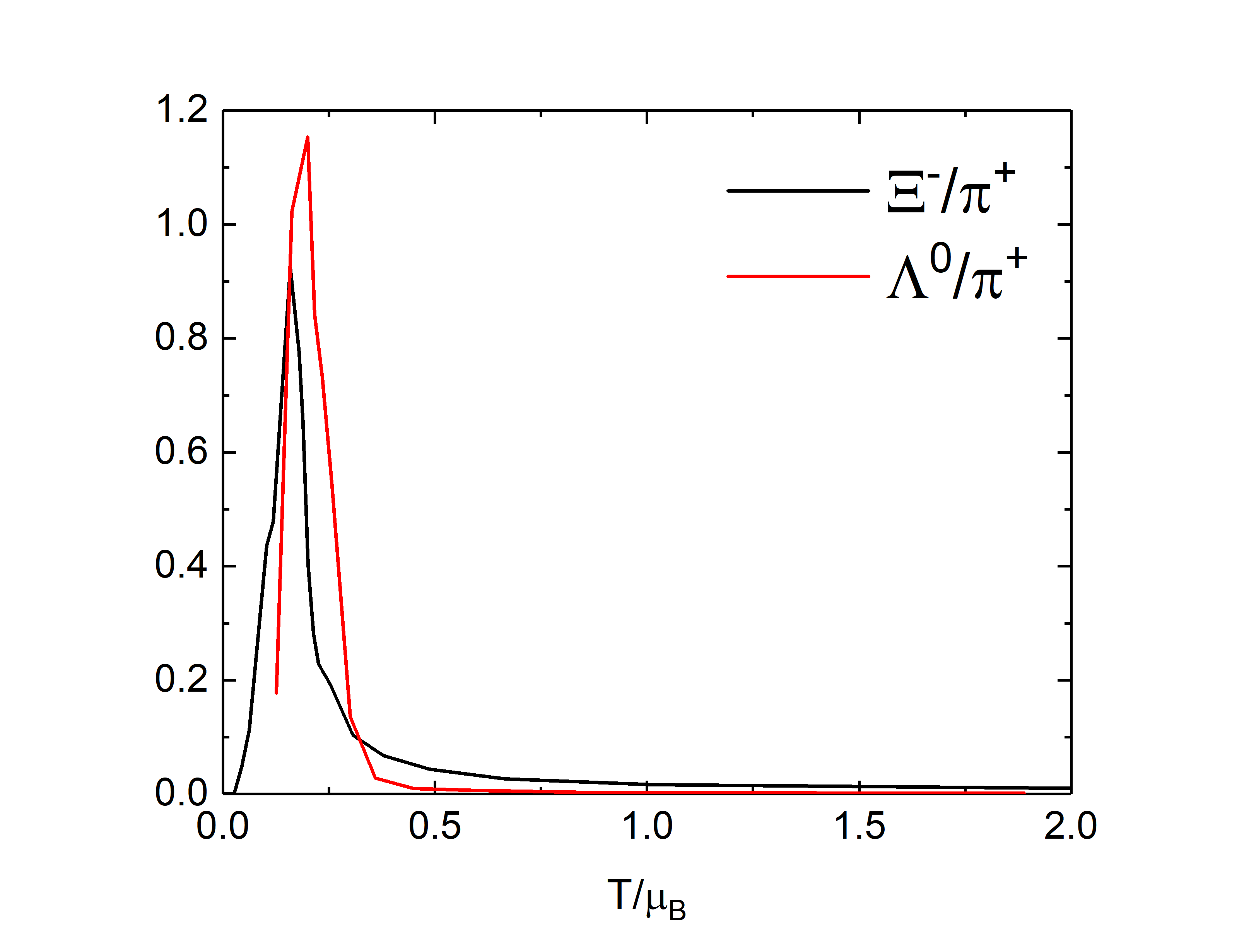}
}
\caption{$\Lambda^0/\pi^+$  (red line) and $\Xi^-/\pi^+$ (black line) ratios.}
\label{bar_pi_ratio}
\end{figure}

The results for baryon-to-meson ratios are presented in Fig.\ref{bar_pi_ratio}. Data for $\Xi^-/\pi^+$ and $\Lambda^0/\pi^+$ were calculated along the lower green and blue curves of phase diagram (rigth panel in Fig.\ref{phase_diagr}) which correspond to $T^{\rm bar}_{\rm dec}$ till 0.373 GeV for $\Xi^-$ and 0.27 GeV for $\Lambda$  and then along the dash-dotted vertical lines, which are supposed now as ''freezeout lines'' at high chemical potential. Both ratios demonstrate the peak-like behaviour.

%%%%%%%%%%%%%%%%%%%%%%%%%%%%%%%%%%%%%%%%%%
\section{Conclusions}

The article summarizes our calculations of the ratios of mesons and baryons with strangeness to nonstrange mesons within the framework of the PNJL-like models. The interest to these ratios appears since they have the ''horn'' structure in their energy dependencies, which is supposed to be a signal of deconfinement and may be sensitive to the structure of the phase diagram, including the position of CEP and TCP \cite{Andronic:2009gj}. Our works show that the $K^+/\pi^+$ ratio is more sensitive to the matter properties, than to the phase diagram structure. This work demonstrates that the EPNJL model reproduces the peak-like structure for $\Lambda^0/\pi^+$ and $\Xi^-/\pi^+$ ratios, but the validity of this estimation is limited by some features of the description of baryons in the model. Therefore, most of our analysis must primarily be taken as qualitative hints, e.g., about the role of the strange
quark chemical potential and pion chemical potential or the effect of the vector interaction.

This work raises several aspects related to the description of baryons as diquark-quark bounded states. The first one is associated with the selection of correct model parameters, which would make it possible to obtain proton and other baryon masses below the threshold  value  $M_D + mq$.  For example, our parameters and choice of the model variation affect the deconfinement temperatures of baryons. The statistical model and the experiment predict a lower chemical freeze-out temperature for proton in comparison with that for $\Xi$. This difference is about 30 MeV \cite{Torres-Rincon:2015rma,Preghenella:2011np}. The PNJL model with our parameters shows 20 MeV, while the  EPNJL model shows 10 MeV.

The second aspect is related to the description of the baryon as a diquark-quark state in the framework of (E)PNJL model. As noted above, this model usually takes into account only scalar part in the mass equations  Eqs.(\ref{BS_bar}) - (\ref{Pol_bar2}), skipping axial-vector part. Nevertheless in works \cite{Barabanov:2020jvn,Cheng:2022jxe} is shown that the accounting for the axial-vector  part in mass equations plays an important role in the correct description of the baryon properties \cite{Cheng:2022jxe}.

The third aspect concerns the description of baryons as a quark-diquark state at a high chemical potential. At low density the two quark pair forms tightly bound localized diquark states   which can pick up another quark with the right colour to form a colour-singlet baryon. The rise of the chemical potential (or density) leads to a weakening of the interaction strength  between quarks and form weakly bounded Cooper pairs in an attractive colour antitriplet channel, leading to the phenomenon of colour superconductivity. However in dense matter the diquark does not have to be stable in order to form a stable baryon, since it can be a bounded state of three quarks, the  so-called Borromean state \cite{Blanquier:2011zz,Wang:2010iu}.

In this situation, further improvements on the more fundamental side, allowing to include the axial-vector part, and describe the baryon above critical densities,  are highly desirable. 

\section{Acknowledgments}

We acknowledge a discussion with D. Blaschke about the baryon description in the frame of the model. We thank A. Radzhabov for his comments on the quasi-chemical potential for pions and strange chemical potential in nonequilibrium systems.

\end{document}